\documentclass[11pt,twoside]{book}
\usepackage{vshalo}
\usepackage{longtable}
\usepackage{amsmath,amssymb}
\usepackage{graphicx}
\usepackage{lscape}
\usepackage{psfig}
\usepackage{epsf}
\usepackage{index}
\usepackage{natbib}
\usepackage{bigdelim}
\usepackage{multirow}
\makeindex
\newindex{aut}{adx}{and}{\Large Author  Index}
\newcommand{\axindex}[1]{\index[aut]{#1}}

\begin{document}

%%%%%%%%%%%%%%%%%%%%%%%%%%%%%%%%%%%%%%%%%%%%%%%%%%%%%%%%%%%%%%%%%%%%%%%%%%
\pagestyle{myheadings}
\setcounter{equation}{0}\setcounter{figure}{0}\setcounter{footnote}{0}\setcounter{section}{0}\setcounter{table}{0} \setcounter{page}{111}

\markboth{Clementini}{RR Lyrae Stars in Dwarf Spheroidal Galaxies}
\title{RR Lyrae Stars in Dwarf Spheroidal Galaxies}
\author{Gisella Clementini}\axindex{Clementini, G.}
\affil{INAF - Osservatorio Astronomico di Bologna, Bologna, Italy}
%
%\pagestyle{myheadings}
%\setcounter{equation}{0}\setcounter{figure}{0}\setcounter{footnote}{0}\setcounter{section}{0}\setcounter{table}{0}\setcounter{page}{35}
%\markboth{Sterken, Samus \& Szabados}{VS-Halo Papers}
%\title{VS-Halo Papers: Instructions}
%\author{Christiaan Sterken$^1$, Nikolay N. Samus$^2$ and Laszlo Szabados$^3$}
%\axindex{Sterken, C.}\axindex{Samus, N. N.}\axindex{Szabados, L.}
%\affil{$^1$Vrije Universiteit Brussel,  Brussels, Belgium\\
%$^2$Institute of Astronomy of the RAS, Moscou, Russia\\
%$^3$Konkoly Observatory of the Hungarian Academy of Sciences, Hungary} 
%%%%%%%%%%%%%%%%%%%%%%%%%%%%%%%%%%%%%%%%%%%%%%%%%%%%%%%%%%%%%%%%%%%%%%%%%%%

\begin{abstract}

With ages comparable to the age of the Universe, the variable stars of RR Lyrae type have eyewitnessed the formation of their host galaxies, and thus 
can provide information on the processes that led to the assembling of large galaxies such as the Milky Way  and the 
Andromeda galaxy. 
The present knowledge of the RR Lyrae population in Local Group dwarf spheroidal galaxies is reviewed,   
calling attention to the ``ultra-faint" spheroidal systems
recently discovered around the Milky Way by the Sloan Digital Sky Survey.
The properties of the RR Lyrae stars and the Oosterhoff dichotomy observed for Galactic globular clusters and field 
RR Lyrae stars are discussed, since they put constrain on the possibility that the Milky Way and Andromeda halos 
were built up from protogalactic fragments resembling the dwarf spheroidals 
we observe today. 
\end{abstract}
\section{Introduction}
In the $\Lambda$-cold-dark-matter theory of galaxy's formation,   
dwarf Spheroidal (dSph)
galaxies are suggested to be the ``building blocks" from which larger galaxies were assembled via hierarchical merging and accretion. 
With their distinctive spheroidal shape,  dSphs are the most common type of galaxies in the local universe. They have high mass-to-light ratios and 
are the most dark matter-dominated objects we know. In the 
Local Group (LG), they are generally found around 
the two large spirals, the Milky Way (MW) and M31.
The MW is surrounded by 10 ``bright" dSph satellites spanning a range from a few to about 250 kpc in distance: Sagittarius, Draco, Fornax, Carina, 
Sculptor, Leo~I, Leo~II, Ursa Minor (UMi), Sextans, and the Canis Major overdensity, whose true nature, whether an actual galaxy 
or the MW warp, is still matter of debate (see, e.g., Mateu et al. 2009, and references therein). Two further ``bright" dSphs, Cetus and Tucana, lie more isolated 
at distances of 780 and 890 kpc, respectively, from the MW.  These systems are generally devoid of gas and host predominantly old stellar populations.
All dSphs show in the color-magnitude diagram (CMD) prominent red giant branches (RGBs), well 
extended horizontal branches (HBs), and  well populated main sequences (MSs). As an example of ``bright" dSph we show in  Fig.1 
the CMD of UMi. 
\begin{figure}[!h]
\centerline{\hbox{\psfig{figure=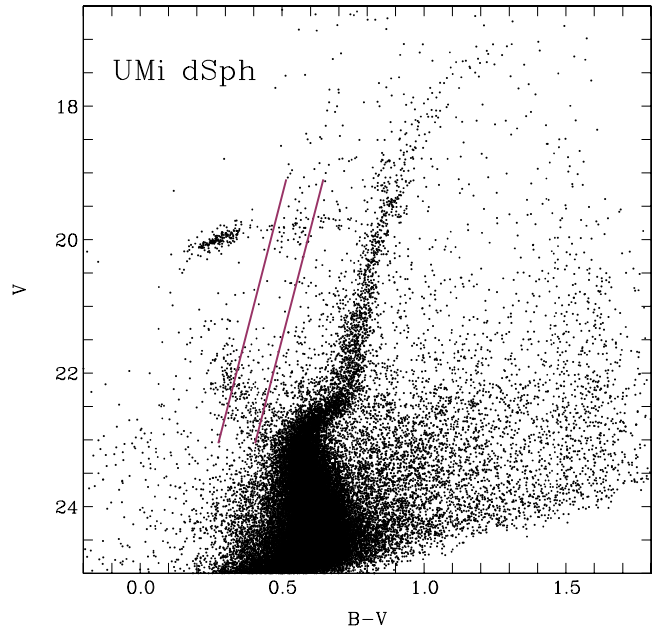,angle=0,clip=,width=9cm}}}
%\centerline{\hbox{\psfig{figure=umi-cmd.eps,angle=0,clip=,width=9cm}}}
\caption[]{The $V, B-V$ color-magnitude diagram of the UMi dSph galaxy, from Dall'Ora et al. (2010). The solid lines show the location of the classical instability
strip for pulsating variable stars. The strip crosses the galaxy horizontal branch where many RR Lyrae stars have been detected, and the blue straggler 
star region, where several variables of SX Phoenicis type were also found.} 
\label{clementini-fig1} 
\end{figure}
The ``bright" dSphs may also contain some younger stars populating the so called ``blue plume", as for instance in Sagittarius (see, e.g., 
Bellazzini et al. 2006), or intermediate-age stars, like in  Fornax (see, e.g., Poretti et al. 2008, and references therein), or show  
complex  star formation histories with multiple bursts of 
star formation, as in Carina (Monelli et al. 2003). The ``bright" dSphs are too few in number, compared to the predictions of the 
$\Lambda$-cold-dark-matter theory, and  contain only very few stars as metal poor as
the stars observed in the  Galactic halo. Since 2005, 15 new dSph satellites were discovered around the MW (see, e.g., Moretti et al. 2009 and references therein, for an 
updated list), mainly based on data collected by the Sloan Digital Sky Survey (SDSS, York et al. 2000).  
The new galaxies are mainly concentrated around the North Galactic pole, as  this is the region  observed by the SDSS, but   
many more are likely to exist at latitudes not explored yet. Thus the census of the MW satellites is likely to increase significantly when 
surveys will extend to cover the full sky. The new dSphs are fainter than the previously known spheroidals, with typical surface 
brightness $\mu_v >$28 mag/arcsec, they are thus named  ``ultra-faint" dSphs. They  have properties intermediate between globular clusters 
(GCs) and dSphs, and contain very metal poor stars,  as metal poor as 
[Fe/H]$\sim -3.0, -4.0$ dex. Often the ``ultra-faint" dSphs  have irregular shape and appear to be distorted by the tidal interaction with the MW.
Like their ``bright" counterparts, they have high mass to light ratios. All ``ultra-faint" dSphs host an ancient population as old as 
$\sim$ 10 Gyr, and have GC-like CMDs, resembling the CMDs of metal poor Galactic clusters like M92, M15 and M68. The CMD of an  ``ultra-faint" 
dSph galaxy, Ursa Major~I (UMa~I), is shown in Fig. 2.  
\begin{figure}[!h]
\centerline{\hbox{\psfig{figure=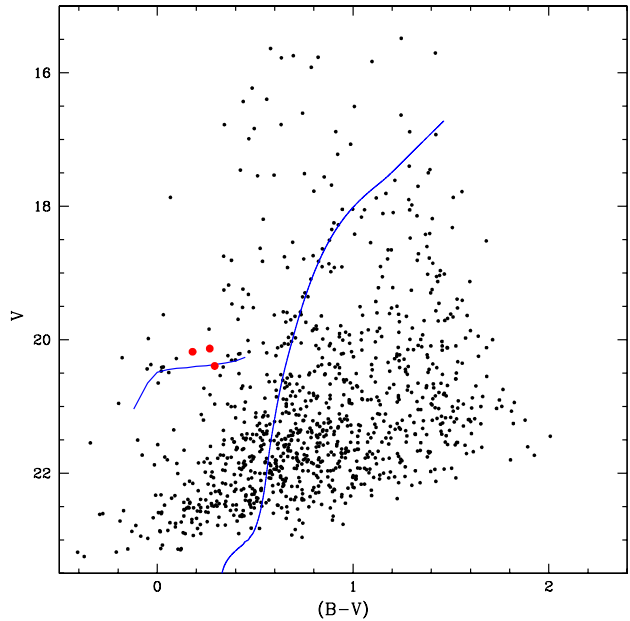,angle=0,clip=,width=9cm}}}
%\centerline{\hbox{\psfig{figure=cmdFin.ps,angle=0,clip=,width=9cm}}}
\caption[]{The $V, B-V$ color magnitude diagram of the UMa~I ``ultra-faint" dSph galaxy, from Garofalo (2009). The solid line is 
the ridge line of the Galactic globular cluster M68 which has been shifted in magnitude and color to match the horizontal branch and the red giant 
branch of UMa~I. Solid circles mark the RR Lyrae stars detected in the galaxy.} 
\label{clementini-fig2} 
\end{figure}

The number of the M31 satellites is poorly constrained.  Twelve dwarf galaxies were known to be M31 companions until 2004,  among which only
6 dSphs, but several new M31 satellites were discovered afterwards. The most recent census is reported by Martin et al. (2009) along with the  
discovery of two new M31 dSph satellites: And XXI and And XXII. As for the MW, the  number of  M31 satellites is likely to increase significantly 
in the near future.

Fig. 3 shows the location of ``bright" and ``ultra-faint" dSphs in 
the absolute magnitude versus half-light radius ($r_h$) diagram. 
The plot is an adapted and updated version of Belokurov et al. (2007) Fig. 8. The MW globulars and some of the M31 GCs are also shown in the figure, 
for a comparison. With their faint luminosities and large dimensions, the ``ultra-faint" dwarfs sample a totally
unexplored region of $M_V - \log (r_h)$ plane. Among the ``ultra-faint" dSphs only Canes Venatici~I (CVn~I), the brightest of these systems, lies 
close to the ``bright" dSphs.
\begin{figure}[!h]
\centerline{\hbox{\psfig{figure=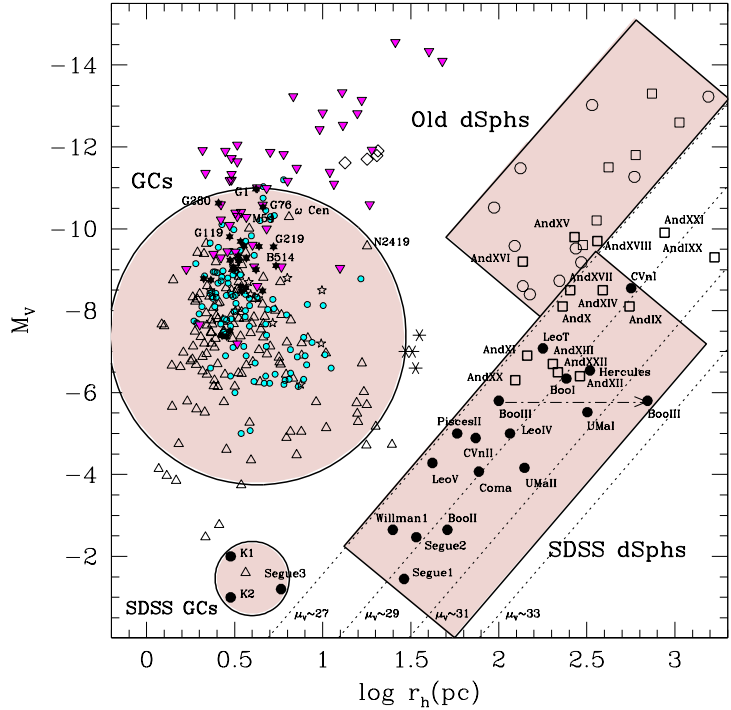,angle=0,clip=,width=11.5cm}}}
%\centerline{\hbox{\psfig{figure=belokurov.eps,angle=0,clip=,width=9cm}}}
\caption[]{Location of ``bright" (``old dSphs" in the figure) and ``ultra-faint" dSphs around the MW and M31 in 
the absolute magnitude versus half-light radius ($r_h$) diagram. Adapted and updated from Belokurov et al. (2007).}
\label{clementini-fig3} 
\end{figure}

\section{Pulsating variable stars in the dSph galaxies} 
The study of the pulsating variable stars offers an additional, very powerful tool   
to get hints on the structure, formation and evolution of the dSph galaxies, and to infer whether they can have contributed to the formation of the 
MW and Andromeda halos. The pulsating variables in dSphs include:  the SX Phoenicis stars, the RR Lyrae stars, the BL Herculis and 
W Virginis variables  
(which are often referred to as Population II Cepheids), and the anomalous Cepheids. Table~1 summarizes their main characteristics. 
 Among them, the RR Lyrae variables are the most common type. These low mass 
(M$< 1 M_\odot$), old ($t \geq 10$ Gyrs) stars can pulsate in the fundamental radial mode (RRab), in the first overtone mode (RRc) or in both modes
simultaneously (RRd). They have been found in ``all" Local Group 
galaxies, irrespective of the galaxy morphological type, thus implying that all nearby galaxies 
started forming stars at an early epoch, just after they were formed.  
\begin{table}[!ht] 
\caption{Main characteristics of different types of pulsating variables found in dSph galaxies.}
\smallskip
\begin{center}
\scriptsize
\begin{tabular}{lccccc}
\tableline
\noalign {\smallskip} 
Type & Period & $M_v$ & Population & Age   & Evolutionary Phase\\ 
     & (days) & (mag) &            & (Gyrs) &  \\
\noalign{\smallskip}
\tableline
\noalign{\smallskip}
SX Phoenicis      & $\leq$ 0.1    &2$\div$3    & II & $>$10 & MS \\
RR Lyrae          & 0.2$\div$1.0  &0$\div$1    & II & $> $10 & HB \\ 
BL Herculis      & $<7 \div 10$  &$-1 \div 0$ & II & $> $10 & post-HB \\ 
W Virginis        & 10$\div$30    &$-3 \div -1$& II & $> $10 & post-HB \\	 
Anomalous Cepheids&0.3$\div$2.5   &$-2 \div0 $ & ?  & $\sim 4 \div 8$ & HB \\ 
\noalign{\smallskip}
\tableline
\noalign{\smallskip}
\end{tabular}  \end{center}
\scriptsize
% $^1$Note: put notes to items here, before the end{table}command. $^2$ Posters may include a Figure  or a Table, if it fits all on 1 page. 
\end{table}
The RR Lyrae stars are also primary distance indicators 
to measure the distance of their host systems, since they obey a luminosity$-$metallicity relation
($M_V (RR) - $[Fe/H]) in the visual band, and a period-luminosity$-$metallicity ($PLZ$) relation in the $K$ band.
They can be used to map the 3D structure and the radial trends occurring  in a galaxy, and to trace halos and streams.
For instance, a local overdensity of RR Lyrae stars in the Galactic halo turned out to be the northern tidal stream left over by the Sagittarius dSph
which is disrupting itself into the MW halo, a clear observational evidence of a merging phenomenon still going on today in the MW halo.
But most importantly, the RR Lyrae stars can provide constrains  on whether the MW was formed by accretion of protogalactic fragments 
resembling the early counterparts of its present-day dSph satellites,  since field and cluster RR Lyrae stars in the MW show well 
distinct properties, that should be also shared by any possible contributor to the MW halo.  In particular, most of the MW GCs 
containing RR Lyrae stars divide into two separate groups according to the mean period of the fundamental-mode RR Lyrae stars  
$\langle Pab \rangle$ and the number of fundamental and first overtone pulsators. This phenomenon is called Oosterhoff dichotomy 
(Oosterhoff 1939). Oosterhoff type I clusters (OoI) have 
$\langle Pab \rangle$ =0.55 days, while Oosterhoff type II clusters (OoI) 
have $\langle Pab \rangle$ =0.64 days (Oosterhoff 1939). The Oosterhoff dichotomy is related to the metallicity, 
since Oo~II clusters are generally 
metal poor with [Fe/H]$\sim -2.0$ dex, while  Oo~I clusters are more metal rich, with  [Fe/H]$\sim -1.4$ dex. In the MW 
Oo~I and Oo~II GCs lie well separated in the $\langle Pab \rangle$ {\it versus} [Fe/H] plane (see Fig. 2 of Smith et al. 2009).  
The Oosterhoff properties of the MW GCs are summarized
in Table~2. 
\begin{table}[!ht] 
\caption{Oosterhoff properties of the MW GCs}
\begin{center}
%\scriptsize
\begin{tabular}{lcccc}
\tableline
\noalign {\smallskip} 
Type & $\langle Pab \rangle$ & $\langle Pc \rangle$ & N$_c$/N$_{total}$ & [Fe/H]\\ 
     & (days) & (days) &            &   \\
\noalign{\smallskip}
\tableline
\noalign{\smallskip}
Oo~I      & 0.55 & 0.32 & 0.17 & $\sim -1.4$ \\
Oo~II      & 0.64 & 0.37 & 0.44 & $\sim -2.0$ \\
\noalign{\smallskip}
\tableline
\noalign{\smallskip}
\end{tabular}  \end{center}
\scriptsize
% $^1$Note: put notes to items here, before the end{table}command. $^2$ Posters may include a Figure  or a Table, if it fits all on 1 page. 
\end{table}

The MW field RR Lyrae stars also seem to show an Oosterhoff dichotomy, as the RR Lyrae stars within the halo belong
predominantly to the Oo~I group, but with a significant Oo~II component appearing to be more concentrated to the Galactic plane
that the Oo~I halo component (see Kinemuchi et al. 2006, Miceli et al. 2008, De Lee 2008, in comparison with Szczygiel et al. 2009).

\subsection{RR Lyrae stars in the MW dSph satellites}

The pulsation properties of the RR Lyrae stars observed in the ``bright" dSphs surrounding the MW are summarized in Table~3.
Only Sagittarius and UMi show distinct Oosterhoff types, while an Oosterhoff dichotomy is not observed among the other ``bright" dSphs, as 
field and cluster RR Lyrae stars in these galaxies %as well as field and clusters RR Lyrae stars
%in the Magellanic Clouds 
have  $\langle Pab \rangle$  intermediate between the two Oosterhoff types and fill the Oosterhoff gap observed in the 
MW (see Fig. 4 in Smith et al. 2009). Thus, the MW is unlikely to have formed by accretion of protogalactic fragments resembling the 
early counterparts of its present-day ``bright" dSph satellites.
\begin{table}[!ht] 
\caption{Oosterhoff properties of RR Lyrae stars in the ``bright" MW dSphs}
\scriptsize
\begin{center}
\scriptsize
\begin{tabular}{lccccc}
\tableline
\noalign {\smallskip} 
Name & Distance$^1$ & $\langle {\rm [Fe/H]} \rangle$$^1$  & N(RRab+c+d)$^1$ & $\langle Pab \rangle$$^1$ & Oo Type$^1$\\ 
     & (kpc)    &                                 &             & (days) &  \\
\noalign{\smallskip}
\tableline
\noalign{\smallskip}
Ursa Minor    &~66$\pm$3 & $-2.2$ & 47+35                & 0.638 & Oo~II \\
Draco         &~82$\pm$6 & $-2.0$ & 214+30+26            & 0.615 & Oo~Int \\
Carina        &101$\pm$5 & $-2.0$ & 54+15+6              & 0.631 & Oo~Int \\ 
Fornax        &138$\pm$8 & $-1.3$ & 396+119($\sim$ 2000) & 0.595 & Oo~Int(field+GCs)\\ 
Sculptor      &~79$\pm$4 & $-1.8$ & 132+74+18:           & 0.587 & Oo~Int \\
Leo~I         &250$\pm$30& $-1.7$ & 47+7($\sim$ 250)     & 0.602 & Oo~Int \\
Leo~II        &205$\pm$12& $-1.9$ & 106+34+8:            & 0.619 & Oo~Int \\
Sextans       &~86$\pm$4 & $-1.7$ & 26+7+3               & 0.606 & Oo~Int \\
              &          &        &                      &       &        \\                   
Sagittarius   &~24$\pm$2 & $-1.55$& $>$4200              & 0.574 & Oo~I(field),I,II,Int(GCs) \\
Canis Major&~~7.1$\pm$0.1&$-1.2/-1.7$&26+17         & 0.57   &  Oo~I/Oo~II(GCs)  \\ 
                      &                           &                   &10+8            &0.56    & \\
                      &                           &                   &30+22         & 0.66   &  \\                   
              &          &        &                      &       &        \\                   
Cetus         &780$\pm$40& $-1.8$ & 147+8+17             & 0.614 & Oo~Int \\
Tucana        &890$\pm$50& $-1.8$ & 216+82+60            & 0.604 & Oo~Int \\
\noalign{\smallskip}
\tableline
\noalign{\smallskip}
\end{tabular}  \end{center}
\scriptsize
$^1$The distance for the Canis Major dSph is from Martin et al. (2004), for Cetus and Tucana from Bernard et al. (2009), for 
all other galaxies from Mateo (1998).\\
$^2$Metal abundance, number of RR Lyrae stars, $\langle Pab \rangle$ values, and Oo types are from the compilation in Smith et al. (2009) 
and references therein, supplemented by data in Bernard et al. (2009) and Catelan (2009).
\end{table}

Table~4 summarizes the properties of the RR Lyrae stars observed in the ``ultra-faint" dSphs. Most of these galaxies contain very few RR Lyrae 
stars. Only Bootes~I and Canes Venatici~I have significant numbers of variables, and can be safely classified respectively as Oo~II and Oo~Int 
systems. 
The Oosterhoff classification of the other ``ultra-faint" dSphs is more insecure given the few variables they contain, nevertheless, 
in so far they can be classified, they all tend to be Oosterhoff type II systems. Thus, it seems that the  MW may have formed
early on by accretion of protogalactic fragments similar to the present-day ``ultra-faint" SDSS dwarfs as they were at earlier times.  

\begin{table}[!ht] 
\caption{Oosterhoff properties of RR Lyrae stars in the ``ultra-faint" MW dSphs}
\scriptsize
\begin{center}
\scriptsize
\begin{tabular}{lccccc}
\tableline
\noalign {\smallskip} 
Name & Distance$^1$ & $\langle {\rm [Fe/H]} \rangle$  & N(RRab+c+d)$^1$ & $\langle Pab \rangle$$^1$ & Oo Type$^1$\\ 
     & (kpc)    &                                 &             & (days) &  \\
\noalign{\smallskip}
\tableline
\noalign{\smallskip}
Bootes~I         &~66 $^{+3}_{-2}$& $-2.5$ & 7+7+1 & 0.69 & Oo~II \\
Canes Venatici~I (CVn~I) &210 $^{+7}_{-5}$& $-2.1$ & 18+5  & 0.60 & Oo~Int \\
Canes Venatici~II (CVn~II)&160 $^{+4}_{-5}$& $-2.3$ & 1+1   & 0.74 & Oo~II? \\ 
Coma Berenices (Coma)   &~42 $^{+2}_{-1}$& $-2.3$ & 1+1   & 0.67 & Oo~II? \\ 
Leo~IV           &154 $\pm$5      & $-2.3$ & 3     & 0.66 & Oo~II? \\
Ursa Major~II (UMa~II) &34.6$\pm$0.7    & $-2.4$ & 1     & 0.66 & Oo~II? \\
Ursa Major~I (UMa~I)   &~95 $\pm$4      & $-2.2$ & 3     & 0.64 & Oo~II? \\
\noalign{\smallskip}
\tableline
\noalign{\smallskip}
\end{tabular}  \end{center}
\scriptsize
$^1$Distances and properties of the RR Lyrae stars are from Dall'Ora et al. (2006) for Bootes~I; from 
Kuehn et al. (2008) for CVn~I; from Greco et al. (2008) for CVn~II; from Musella et al. (2009) for Coma,
from Moretti et al. (2009) for Leo~IV; from Dall'Ora et al. (2009) for UMa~II; and from Garofalo (2009) for
UMa~I. Question marks indicate the difficulty of classifying these galaxies due the small number of variables they contain.
\end{table}

\subsection{RR Lyrae stars in the M31 satellites}

Very little is known about the existence of an Oosterhoff dichotomy in M31, since only recently studies 
providing a characterization of the RR Lyrae stars in the M31 field and globular clusters
have commenced to appear in the literature, thanks to the collection of Hubble Space Telescope (HST) observations.
Twenty-nine RRab's, 25 RRc's, and 1 RRd were discovered in an ACS@HST field 11 kpc from the center of M31 by Brown et al (2004). 
According to their $\langle Pab \rangle$ = 0.594 days, and $\langle Pc \rangle$ = 0.353 days values Brown et al. concluded that the 
M31 field is Oo-Intermediate. A different conclusion was reached instead by Sarajedini et al. (2009), who obtained  
$\langle Pab \rangle$ = 0.557 days, and $\langle Pc \rangle$ = 0.327 days, from 555 RRab's and 126 RRc's in two ACS fields near M32, 
thus concluding that these M31 fields have Oo~I properties.  
RR Lyrae stars in 4 M31 GCs were detected by Clementini et al. (2001) using HST archival data, but no reliable period derivation was possible. More
recently, Clementini et al. (2009) identified and derived periods for a large number of RR  Lyrae stars in B514, one of the brightest 
($M_V \sim -9.1$ mag) and largest  ($r_h \sim 5.4$ pc) metal poor ([Fe/H]$\sim -2.0$ dex) clusters in M31, 
about 55 kpc from the galaxy center. They found 82 RRab's and 7 RRc's having $\langle Pab \rangle$ = 0.58 days and $\langle Pc \rangle$ = 0.35 
days, and concluded that in spite of the low metal abundance B514 appears to be a borderline Oo~I cluster. 

RR Lyrae stars have been studied in 5 of the dSphs surrounding M31. Results for these galaxies are summarized in Table 5. 
Detection of RR Lyrae stars, without reliable period determinations, is also reported  for three of the big M31 dSphs, namely, NGC147, NGC185 and 
NGC205, which were found to contain respectively 36 
(Yang \& Sarajedini 2010, and references therein), 151 (Saha et al. 1990),  and 30 (Saha et al. 1992)
candidate RR Lyrae stars. 
The studies of the M31 variables seem to indicate that the Andromeda RR Lyrae stars may have slightly different properties than the MW variables.

\begin{table}[!ht] 
\caption{Oosterhoff properties of RR Lyrae stars in the M31 dSphs}
%\scriptsize
\begin{center}
%\scriptsize
\begin{tabular}{lccccc}
\tableline
\noalign {\smallskip} 
Name &Distance$^1$  & $\langle {\rm [Fe/H]} \rangle$  & N(RRab+c+d)$^1$ & $\langle Pab \rangle$$^1$ & Oo Type$^1$\\ 
     & (kpc)        &                                 &                 &       (days)              &            \\
\noalign{\smallskip}
\tableline
\noalign{\smallskip}
And~I      &765$\pm$25&    $-1.5$ & 72+26 & 0.575 & Oo~I/Int \\
And~II     &665$\pm$20&    $-1.5$ & 64+8  & 0.571 & Oo~I \\
And~III    &740$\pm$20&    $-1.9$ & 39+12 & 0.657 & Oo~II \\
And~V      &820$\pm$16&    $-2.2$ & ~7+3  & 0.685 & Oo~II  \\
And~VI     &815$\pm$25&    $-1.6$ & 91+20 & 0.588 & Oo~Int \\
\noalign{\smallskip}
\tableline
\noalign{\smallskip}
\end{tabular}  \end{center}
%\scriptsize
$^1$Distances and properties of the RR Lyrae stars are from Pritzl et al. (2002) for And~VI; from  Pritzl et al. (2004) for 
And~II; from  Pritzl et al. (2005) for And~I and III; and from Mancone \& Sarajedini (2008) and Sarajedini (2009) for And~V.
\end{table}

\section{Summary}
Most of the RRab stars in the MW and M31 halos have Oosterhoff type I properties, but the inner MW halo also contains a significant Oo~II component.
By contrast the Oo~I type is rare among the dSph galaxies, where only Sagittarius is Oo~I. The ``bright" MW dSphs tend to be Oo-Intermediate, while 
the ``ultra-faint" dSphs tend to have Oo~II type, in so far they can be classified.
The Galaxy is unlike to have formed by accretion of protogalactic fragments
resembling its present-day ``bright" dwarf satellites. On the other hand, systems resembling the ``ultra-faint" dSphs may have contributed in the past to the formation 
of the MW halo. 
The study of the M31 RR Lyrae population is still in pioneering stage. Further data for both field and 
cluster variables are needed to reach firm conclusions on the Oosterhoff classification of the Andromeda galaxy.

%\newpage
%\section{Example of Table}

%\section{Examples of citing in text}

%You may write Cousins (2001),  Crawford \& Barnes (1970) or use \begin{verbatim}\citet{Cousins2001}, \citep{Cousins2001}, \citet{Crawford&Barnes1970}\end{verbatim}

%\noindent Please make sure that every cite corresponds to a reference, and vice versa.

%\section{Example of References}

\end{document}